\newcommand{\inmaindoc}{} 

\newcommand{\arXiv}{} 

\ifdefined\anon
\newcommand{\anonsuf}{-anon} 
\else
\newcommand{\anonsuf}{} 
\fi

\ifdefined\nips
\documentclass{article}

\usepackage[final]{nips_2016}

\else
\documentclass[10pt, conference, compsocconf]{IEEEtran}
\fi

\def\vsp{\vspace{-0.05in}}
\usepackage[utf8]{inputenc} 
\usepackage[T1]{fontenc}    
\ifdefined\arXiv
\usepackage{hyperref}       
\else
\usepackage[draft]{hyperref}       
\fi

\usepackage{url}            
\usepackage{booktabs}       
\usepackage{amsfonts}       
\usepackage{nicefrac}       
\usepackage{microtype}      

\usepackage[pdftex]{graphicx}

\usepackage{mathtools}
\usepackage{listings}
\DeclarePairedDelimiter\ceil{\lceil}{\rceil}
\DeclarePairedDelimiter\floor{\lfloor}{\rfloor}
\newcommand{\dx}{{\times}}

\ifdefined\anon
\newcommand{\bodartc}{}
\newcommand{\boda}{Framework}
\else
\newcommand{\bodartc}{Boda-RTC: }
\newcommand{\boda}{Boda}
\fi
\title{\bodartc~Productive Generation of Portable, Efficient Code for Convolutional Neural Networks on Mobile Computing Platforms}

\ifdefined\anon
\author{ Authors withheld for blind review. }
\else
\author{
  Matthew W. Moskewicz \qquad Forrest N. Iandola \qquad Kurt Keutzer\\
  University of California, Berkeley\\
  \texttt{\{moskewcz,forresti,keutzer\}@eecs.berkeley.edu}\\
}
\fi

\begin{document}
\newcommand{\ndaii}[4]{\ensuremath{#1 \dx #2 = #3 \dx #4}}
\newcommand{\ndaiii}[6]{\ensuremath{#1 \dx #2 \dx #3 = #4 \dx #5 \dx #6}}
\newcommand{\ndaiiii}[8]{\ensuremath{#1 \dx #2 \dx #3 \dx #4 = #5 \dx #6 \dx #7 \dx #8}}


\maketitle

\ifdefined\arXiv
\thispagestyle{plain}
\pagestyle{plain}
\fi

\begin{abstract}
The popularity of neural networks (NNs) spans academia~\cite{schmidhuber2015deep}, industry~\cite{amodei2015deep}, and popular culture~\cite{silver2016mastering}.
In particular, convolutional neural networks (CNNs) have been applied to many image based machine learning tasks and have yielded strong results~\cite{razavian2014cnn}.
The availability of hardware/software systems for efficient training and deployment of large and/or deep CNN models is critical for the continued success of the field~\cite{oh2004gpu}~\cite{schmidhuber2015deep}.
Early systems for NN computation focused on leveraging existing dense linear algebra techniques and libraries~\cite{alexnet}~\cite{jia2014caffe}.
Current approaches use low-level machine specific programming~\cite{maxDNN} and/or closed-source, purpose-built vendor libraries~\cite{cuDNN}.
In this work, we present an open source system that, compared to existing approaches, achieves competitive computational speed while achieving significantly greater portability.
We achieve this by targeting the vendor-neutral OpenCL platform~\cite{stone2010opencl} using a code-generation approach.
We argue that our approach allows for both:
(1) the rapid development of new computational kernels for existing hardware targets, and
(2) the rapid tuning of existing computational kernels for new hardware targets.
Results are presented for a case study of targeting the Qualcomm Snapdragon 820 mobile computing platform~\cite{SD820} for CNN deployment.

\end{abstract}

\begin{IEEEkeywords}
computer vision; code generation; neural networks; mobile computing; convolution
\end{IEEEkeywords}

\section{Introduction and Motivation}
\label{sec:intro}
\vsp

\begin{figure}[ht]
  \centering
  \includegraphics[width=0.8\linewidth]{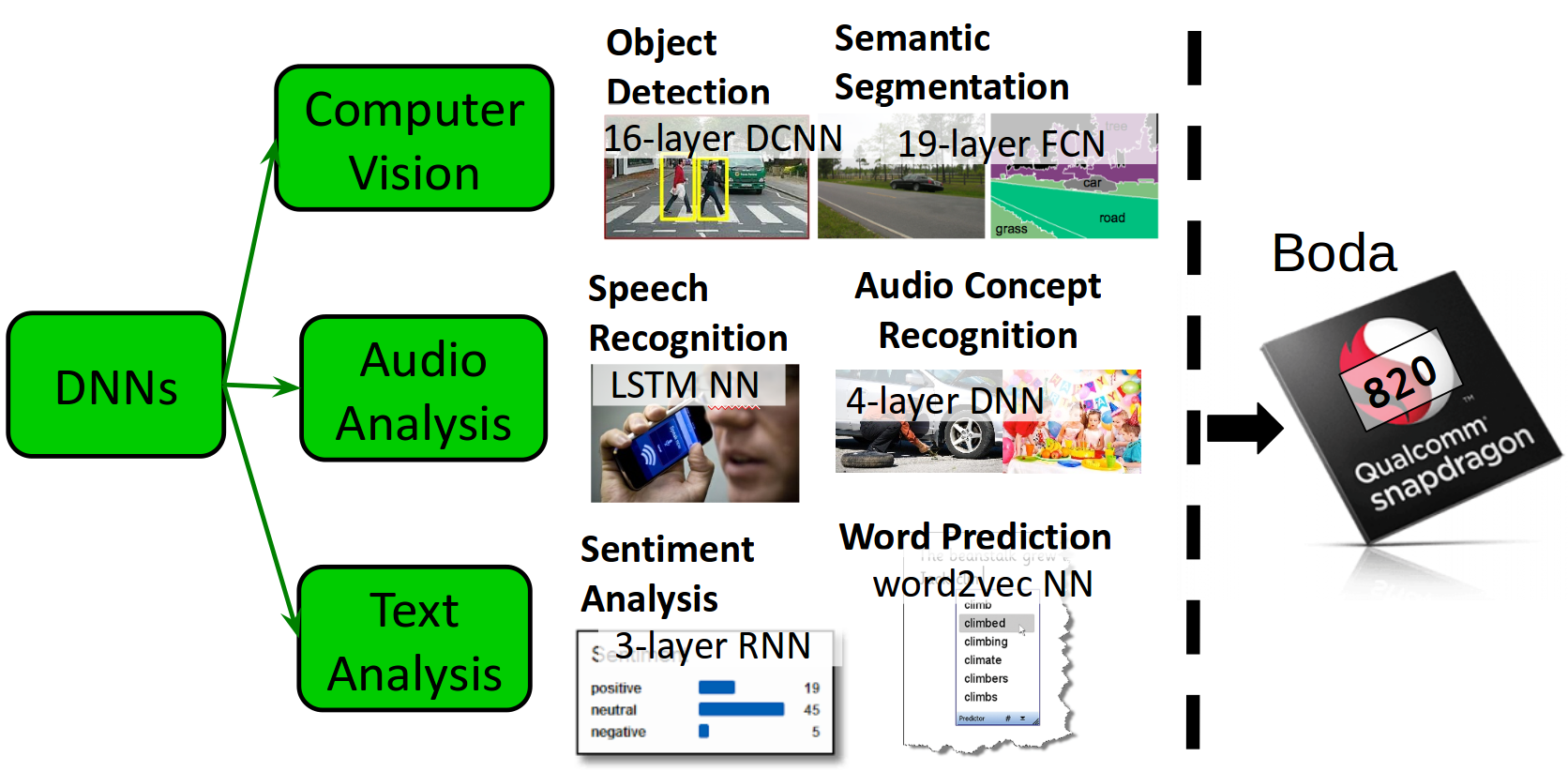}
  \caption{\boda~overview: from NNs to mobile computing platforms.}
  \label{fig:boda-overview}
\end{figure}

Convolutional neural network (CNN) based approaches have become dominant in a broad variety of computer vision applications, including object detection~\cite{girshick2015deformable}, video classification~\cite{karpathy2014large}, and human action recognition~\cite{ji20133d}.
On a mobile phone or wearable device, energy-efficient computer vision is necessary to put these and other applications into production in order to enable novel functionality and achieve the promises of augmented reality.
More critically, the safety of autonomous vehicles hinges on their ability to understand their surroundings in real-time under tight energy, power, and price constraints.
To support ongoing research, development, and deployment of systems that include NNs, it is desirable to nurture a diverse enabling ecosystem of tools and approaches.
In particular, we feel it is desirable to support many hardware and software platforms to enable new applications across many areas, including mobile, IoT, transportation, medical, and others.
Imagine that, for a given task, high-performance vendor libraries exist for at least one platform.
Currently, for CNNs, the vendor is Nvidia, the platform is Maxwell, and the library is cuDNN~\cite{cuDNN}.
Why not simply use that vendor's platform and libraries for the task and be satisfied?
One issue is quite simple: in industrial use cases, choice of platform may be dictated by business concerns.
Further, those same business concerns may preclude dependence on any single vendor.
For example, the flagship Samsung Galaxy S7 mobile phone ships in two versions: one using a Samsung-proprietary Exynos 8890 System-on-Chip (SoC), the other using the Qualcomm Snapdragon 820~\cite{SD820} SoC.
Neither of these SoCs contains Nvidia GPUs or are otherwise capable of running cuDNN.
Further, Nvidia, Qualcomm, and Samsung have engaged in a long running patent dispute over GPU technologies.
Based on the uncertainties associated with such litigation, SoC and/or GPU alternatives are subject to constant change.
Further, even once a hardware platform is chosen, business needs may dictate the specific software tasks that must be supported.
For example, in an effort to provide differentiation, vendors have a strong desire to implement novel functionality not present in existing publicly available libraries.
Together, these uncertainties about both target hardware and particular use-case create a strong pressure for \textit{portability}: the ability to quickly achieve reasonably performance for a variety of tasks across a variety of platforms.

This work focuses on the challenges associated with achieving portable, efficient computation of the key primitive operations needed by convolutional neural networks. We frame the problem as follows:

Consider a use-case consisting of a specific combination of:
\begin{itemize}
\item a target computational device (i.e. a hardware architecture), and
\item a set of convolutional neural network primitives and an input size (i.e. a CNN architecture and task).
\end{itemize}
For such a use-case, achieving good computational efficiency and/or meeting particular performance requirements is, in general, difficult.
Building systems based on existing computational libraries (e.g. BLAS) generally achieves only limited efficiencies~\cite{soumith-bench}.
Improving on such approaches requires tuning multiple computational kernels for the particular use-case at hand.
This generally requires months of effort by a very specialized programmer.
This programmer must be one that is both capable of producing high-efficiency code for a target platform as well as being familiar with the details of CNN computations.
Such programmers are not common and thus their time is a very limited resource.

In this work, we present an open-source vertically-integrated framework for developing and deploying CNN computations.
The framework combines parts of the functionality of CNN middleware frameworks such as Caffe~\cite{jia2014caffe} or TensorFlow~\cite{tensorflow2015-whitepaper} with the functionality of CNN computation libraries such as Nvidia's cuDNN~\cite{cuDNN} or Nervana System's NervanaGPU/neon~\cite{nervanagpu}~\cite{neon}.
``Out-of-the-box,'' our framework does not attempt to have the breadth or depth of features of a typical general-use middleware such as Caffe or TensorFlow.
Also, it does not attempt to achieve the same efficiency as a highly-target-specific computational library such as cuDNN.
Instead, we aim to allow for the rapid development and deployment of new use-cases/flows of the form described above.
In particular, the framework enables productive, flexible, modular code generation of the full set of needed CNN operations for any given use case.

This allows for both:
\begin{itemize}
\item the rapid development of new computational kernels for existing hardware targets, and
\item the rapid tuning of existing computational kernels for new hardware targets.
\end{itemize}

To provide support for these claims, we present a case study of using the framework to target the Qualcomm Snapdragon 820 mobile computing platform~\cite{SD820} for DNN/CNN deployment (see Figure~\ref{fig:boda-overview}).
We show that the framework enabled a quick path to initial functional correctness, and from there helped navigate a smooth path to reasonably efficient computational performance.


The rest of the paper is organized as follows.
In Section~\ref{sec:intro-cnn-ops} we review the semantics of the key computational operations needed to support CNNs, focusing particularly on convolution.
Then, in Section~\ref{sec:conv-calc} we discuss performing CNN calculations on modern computational hardware, focusing particularly on the issue of data reuse.
We take a historical approach to this explanation, citing related work along the way.
In Section~\ref{sec:progport}, we consider the general topic of programming GPUs and performance portability.
In Section~\ref{sec:sgemm} we introduce our framework using the example kernel of single-precision matrix-matrix multiply (SGEMM).
The use of such a relatively simple, existing computational kernel allows us to:
\begin{itemize}
\item illustrate in general what our framework does and how to apply it,
\item to show in particular how we approach a new hardware target, and
\item to provide a good baseline for general performance evaluation.
\end{itemize}
Then, we present our core contribution of productive, portable, efficient, code generation for CNN operations (via case study) in Sections~\ref{sec:gencnnconv} and~\ref{sec:results}.
Finally, we conclude in Section~\ref{sec:conclusions}.

\section{Introduction to CNN Computations}
\label{sec:intro-cnn-ops}
\vsp

Different CNN architectures can contain a wide variety of computational primitives.
For completeness, it is important that any CNN computation system supports a reasonably broad set of such primitives.
Further, it is desirable that support for new operations, particularly those that are simple, can be easily added.
The most common operations needed for deployment (also known as testing or inference, and as opposed to training) include \textit{convolution}, \textit{pooling} and \textit{softmax}.
However, across many common networks, such as AlexNet~\cite{alexnet}, GoogLeNetV1~\cite{googlenet}, and the VGG networks~\cite{simonyan2014very}, \textit{convolution} operations dominate the overall computation.
Thus, in this section, we focus on the convolution operation.
In particular, we consider the commonly occurring forms of convolutions from the above (and similar) CNNs.
Given the nature of this work, we will focus on practical, operational definitions of the relevant computations and the data on which those computations act.
Although various types of data are used in CNNs, the most common are 32-bit (and, increasingly, 16-bit) floating point numbers.
Regardless of exact type or precision, the numeric values used by CNNs are grouped into named ND-Arrays (i.e. collections of numbers with N indexes, sometimes also called N-D Matrices or tensors).
Thus, we define convolution operationally as the function $out = conv(in,filts)$ where $out$, $in$, and $filts$ are all N-D arrays.
For simplicity, we omit discussion of the common practice of special-case handling of biases here.
Further, we will restrict the discussion to convolution over 2D images, which is the only type used in the above example networks.
Thus, for a single 2D multi-channel image, both $in$ and $out$ are 3D arrays, with their dimensions being the image height, the image width, and the number of image channels.
For example, the overall input to a network might be an RGB (3 channel) image with a size of $205 \dx 205$ pixels.
Thus the dimensions of the 3D array storing this image ($dims(in)$) could be written compactly as \ndaiii{Y}{X}{C}{205}{205}{3}, where the Y, X, and C \textit{name} the dimensions, and the 205, 205, and 3 are the \textit{concrete sizes} of those dimensions.
Often, particularly for training, it is desirable to process a \textit{batch} of multiple images.
In this case, an additional $B$ dimension is added to $in$ and $out$.
For the example of a 32 image batch, $dims(in)$ becomes \ndaiiii{B}{Y}{X}{C}{32}{205}{205}{3}.
The results of the convolution are fully independent across input/output image pairs.
Thus, for simplicity, we assume a single image for the remainder of this discussion.
The semantics of convolutions are determined by several architecturally specified values: the number of output channels, the kernel size, and the stride.
While in general the kernel size and stride can be different in each spatial dimension (i.e. X and Y in the 2D case), for simplicity, here we consider only the case where kernel size and stride are the same for all dimensions (i.e. are scalars).
The kernel size and number of output channels, combined with the number of channels in $in$, determine $dims(filts)$.
In our running example, recall that the number of input channels $IC=3$.
If the number of output channels $OC=96$, and the kernel size $KSZ=7$, then $dims(filts)$ will be \ndaiiii{OC}{KSZ}{KSZ}{IC}{96}{7}{7}{3}.
Finally, the width of the output will be given by $out_X = 1 + (in_X - KSZ) / stride$ (and similarly for the height).
Thus, if the stride for this example is 2, then we have $out_X = 1+(205-7)/2 = 100$, and $dims(out)$ will be \ndaiii{Y}{X}{C}{100}{100}{96}.
Note that $in$ is often padded by $\floor{KSZ/2}$ elements (usually zeros) in each spatial dimension.
When such padding is applied (assuming $stride=1$), $out$ will have the same spatial dimensions (height and width) as $in$.
We can calculate each output channel $oc$ using $in$ and the 3D slice of $filts$ where $OC=oc$.
That is, we use each slice $filts[OC=oc]$ (dims \ndaiii{KSZ}{KSZ}{IC}{7}{7}{3}) to compute each slice $out[OC=oc]$ (dims \ndaii{Y}{X}{100}{100}).
Then, for each output point $out[OC=oc,Y=oy,X=ox]$, we extract a $KSZ \dx KSZ$ window of $in$ (sometimes called an \textit{input patch} or \textit{patch}) starting at $Y=oy*stride, X=ox*stride$, across all input channels, to yield the slice $in[Y=[oy*stride,oy*stride+KSZ),X=[ox*stride,ox*stride+KSZ)]$ (dims \ndaiii{Y}{X}{C}{7}{7}{3}).
The final value of each output point is the sum of all elements of the element-wise product of the per-output-channel slice of $filts$ and the per-output-x-y-point slice of $in$.
Typically some \textit{activation function}, such as ReLU ($max(0,x)$)~\cite{ReLU}, is next applied to each output value, and this operation is often fused into the convolution operation itself for efficiency.
For later reference, note here that each $in$ slice is reused across all output channels, and each $filts$ slice is reused across all output x-y points.  
Similarly, note that if multiple images are processed in a batch, the number of input-patches/output-x-y-points is multiplied by $B$.


\section{CNN Computations on Modern Hardware And Related Work}
\label{sec:conv-calc}
\vsp

In the prior section, the final computation of each (scalar) output value in a convolution consisted of an element-wise product of two ND-Arrays (one a slice of the input, one a slice of the filters, with exactly equal dimensions), followed by a sum over all the elements of the product.
If we were to conceptually \textit{reshape} or view the two 3D slices as 1D arrays (i.e. vectors, with size equal to the product of the 3 3D-Array dimension sizes), we can see that this operation is simply a vector dot-product.
Thus, the core computational operation of a large class of NNs (including CNNs) is to perform many dot-product operations between a large set of input vectors and model parameter vectors (also known as \textit{filter weights}, \textit{filters}, or \textit{weights}).
In NNs with multiple \textit{layers}, the outputs of one operation may become inputs to another.
In CNNs, the slices that form the inputs to each dot-product may spatially overlap and thus share data with each other.
Recall that in our running example, the we have $dims(out)$ of \ndaiii{Y}{X}{C}{100}{100}{96}, yielding $100*100*96 = 960000$ output points, each requiring one dot-product to compute.
If we consider the set of all input-slice/filter-slice pairs to these $960000$ dot-products, we see that it is formed from the cross product of $100*100 = 10000$ input slices and $96$ (per-output channel) filter slices.
Thus, each dot-product input is reused across many dot-products: $96$ dot-products for each input-slice, and 10000 dot-products for each filter-slice.

At least as early as 2004, work using GPUs to accelerate NNs made the key observation that this data reuse pattern of NNs makes then well suited for calculation using GPUs~\cite{oh2004gpu}.
Based on the amount of calculation hardware available, and the rate at which it can perform calculations, each computational system has a \textit{peak} computational rate.
Computations which can achieve this peak rate on a given system are termed \textit{compute limited}.
But, on modern computational systems (CPUs and GPUs), computations such as batches of independent dot-product operations are instead limited by the time taken to transfer operands between different storage locations in the system.
This is termed a \textit{bandwidth limited} computation.
Depending on the ratio of communication resources to compute resources in a system, it is necessary to reuse data to varying degrees at different levels of the storage hierarchy of the system to avoid becoming bandwidth limited~\cite{williams2009roofline}.
Over time, the relative cost of communication has increased compared to that of computation, so systems that have high absolute peak computational rates (for a given power level) require increasing amounts of data reuse to achieve those rates~\cite{demmel2013communication}.
The terms \textit{Arithmetic Intensity} (or sometimes more generally \textit{Operational Intensity}), abbreviated \textit{AI} in this work, is used to refer ratios between an amount of computation and an amount of communication.
The most common units for AI are $FLOPS/byte$ ($F/B$), and care must be taken to properly convert any quantities of communication from elements (e.g. 32-bit or 16-bit floats) into bytes (e.g. 4 bytes per 32-bit float, 2 bytes per 16-bit float).
For a given flow, there are two key top-level AI values of interest: the AI required by the hardware to achieve peak compute rate (the \textit{knee} AI), and the best-case AI available in the computation to be performed.
To find the top-level knee AI for a given hardware device, we simply divide the peak compute rate of the device by the peak off-chip bandwidth of the device.
However, note that since most hardware systems have multiple levels of memory hierarchy and have many quirks that complicate issues, this top-level knee AI provides only (perhaps unachievable) upper bounds on performance with respect to any given computation's AI.
To calculate the best-case available AI for a given computation, we take the number of FLOPS to be performed divided by the (minimal) total number of bytes to transfer.

Returning to our working example, recall that the size of each input-slice is $7 \dx 7 \dx 3$, or $147$ elements.
If we form a 2D matrix using all $10000$ input slices as rows, we will have a $10000 \dx 147$ matrix $inmat$.
Similarly, we can form 2D matrix using all $96$ filter slices as columns, yielding a $147 \dx 96$ matrix $filtsmat$.
We can now express the entire convolution operation as a single matrix-matrix multiply operation: $outmat = inmat*filtsmat$.
Note that $outmat$ has $10000*96 = 960000$ elements, so we can reshape it to the desired $dims(out)$ of \ndaiii{Y}{X}{C}{100}{100}{96}.
Due to the dot-product-input reuse discussed earlier, such matrix-matrix multiplies are generally amenable to high-efficiency GPU implementations~\cite{volkov2008benchmarking}.
Quantitatively, this is seen by the fact that such computations have high AI.
\newcommand{\matofsize}[3]{
\begin{bmatrix}
#1_{1,1} & #1_{1,2} & \ldots & #1_{1,#3} \\
#1_{2,1} & #1_{2,2} & \ldots & #1_{2,#3} \\
\vdots & \vdots & \ddots & \vdots \\
#1_{#2,1} & #1_{#2,2} & \ldots & #1_{#2,#3}
\end{bmatrix}
}

\begin{figure*}[t]
\begin{equation*}
  \frac{\matofsize{o}{10000}{96}}{outmat: 96*10K*4B=3.84MB} = \frac{\matofsize{i}{10000}{147}}{inmat: 10K*147*4B=5.88MB} \dx
  \frac{\matofsize{f}{147}{96}}{filtsmat: 147*96*4B=0.06MB}
\end{equation*}  
\begin{equation*}
  FLOPS = 10K*96*147=141GF; Bytes = 3.84+5.88+0.06=9.78MB \rightarrow AI = 141MF/9.78MB = 14.4 F/B
\end{equation*}  
\caption{Example calculation of data reuse / Arithmetic Intensity (AI) in Matrix-Matrix Multiplication}
\label{fig:matmul-ai}
\end{figure*}

In Figure~\ref{fig:matmul-ai}, we show the calculation of the AI for the matrix-multiplication of our running example.

Early CNN frameworks such as cuda-convnet~\cite{alexnet} and Caffe~\cite{jia2014caffe} originally performed CNN convolutions in exactly this fashion, leveraging Nvidia's cuBLAS~\cite{cuBLAS} matrix library.

However, there are several limitations of this BLAS-based approach:
\begin{itemize}
\item When $in_X, in_Y >> KSZ$, $inmat$ is roughly $(KSZ/stride)^2$ larger than $in$; thus explicitly creating the matrix $inmat$ may require significant intermediate memory, and to a lesser extent, non-negligible time.
\item It does not allow for data reuse between spatially overlapping input slices.
\item The underlying matrix-matrix multiply library may not be well optimized for the problem sizes required.
\item It is not possible to fuse an activation function with the convolution operation.
\item It is not possible to use various other optimizations, such as Winograd convolution~\cite{lavin2015fast}.
\end{itemize}
Regardless of the exact reasons, it became clear to the community of researchers working on high-performance implementations of CNNs that BLAS-based approaches were often falling far short of peak compute.
Overall community efforts then turned to purpose-built libraries for convolution, so it remains an open research question what the limits of BLAS-based approaches are in various cases.

In particular, Nvidia soon released the cuDNN~\cite{cuDNN} library, which exposes an API for directly performing convolutions using a variety of approaches.
Due to the closed-source nature of the library, and the fact that it has evolved rapidly, it is difficult to analyze in detail.
However, from community benchmarking it is clear that it achieves much higher efficiency than BLAS-based approaches~\cite{soumith-bench}.
Concurrently, a family of libraries based on an assembly-language-level programming flow appeared~\cite{maxDNN}~\cite{nervanagpu}~\cite{neon}, offering similar performance to cuDNN.
Historically, the assembly language level approach has offered the best performance at any given time, with cuDNN catching up in its next release.
Since the above mentioned assembly-language kernels are open source, one could speculate that Nvidia is copying or reimplementing them internally.
If true, it would imply that the entire high-performance CNN computation ecosystem is anchored by only a few high-performance programmers at Nervana Systems and Nvidia.
One result of this state of affairs is that there is effectively only one usable hardware platform for high efficiency CNN computations: Nvidia Maxwell GPUs.
For completeness, we briefly consider the state of the art for performing convolution on CPUs.
Recently, Intel has made significant advances in supporting convolutions, and is competitive with GPUs at large scale~\cite{das2016distributed}.
However, in practice, most use cases are still single socket, or at most single node, and current CPUs cannot offer competitive performance at this scale, particularly using consumer (as opposed to server) parts.
Also, it is unclear that CPUs are currently competitive with GPUs on a power or cost basis for CNN computation.
Note that a detailed comparison of CPUs vs. GPUs is outside the scope of this paper.
While in this work we focus on targeting GPUs, extending our results to CPUs is a reasonable topic for future work.

\section{Programming GPUs and Performance Portability}
\label{sec:progport}
\vsp
GPUs are generally considered more difficult to program than CPUs.
This is not an issue for an end-user of a library like cuDNN, since all interactions with the GPU are hidden behind a C library interface.
However, if one wishes to \textit{write} such a library, one must face the complexity and issues of GPU programming and portability.
For Nvidia hardware, the proprietary CUDA language is the officially blessed programming language.
Alternately, the OpenCL~\cite{stone2010opencl} standard is the only portable option for targeting a variety of GPUs (including Nvidia GPUs).
In terms of low-level features, memory model, and threading model, CUDA and OpenCL are quite similar.
For example, many GPUs allow explicit loads and stores to the L2 cache memory as an alternative or adjunct to traditional caching and prefetching.
OpenCL and CUDA both expose this; OpenCL calls it \textit{local} memory whereas CUDA calls it \textit{shared} memory.
For the program itself, both languages are based on defining a single function (termed a \textit{kernel} in both languages) that is run by many threads in parallel.
Kernels are compiled to sequences of machine code and cached in instruction caches on GPUs similarly to on CPUs.
In practice, when running the same program on \textit{different} GPUs, OpenCL provides only \textit{functional} portability.
That is, a given (correct) OpenCL program will produce consistent results on all supported OpenCL platforms.
However, an OpenCL program that is tuned for high efficiency on one platform (e.g. an Nvidia Maxwell GPU) will not necessarily deliver high efficiency when run on a different platform (e.g. a Qualcomm Snapdragon SoC's GPU).
Perhaps due to this issue, no OpenCL library comparable to cuDNN or NervanaGPU exists.
Thus, OpenCL support for CNN operations on AMD or any other GPU platforms is generally lacking.
Currently, in the caffe framework, support for OpenCL uses only the BLAS-based approach, and even that support is marginalized and fragmented across various forks~\cite{openclcaffe} and pull requests.
In this paper, we aim to bridge the performance-portability gap for CNN operations, and bring such operations to a more even footing across various hardware platforms when compared with existing high efficiency approaches.

\begin{figure}[ht]
  \centering
  \includegraphics[width=1.0\linewidth]{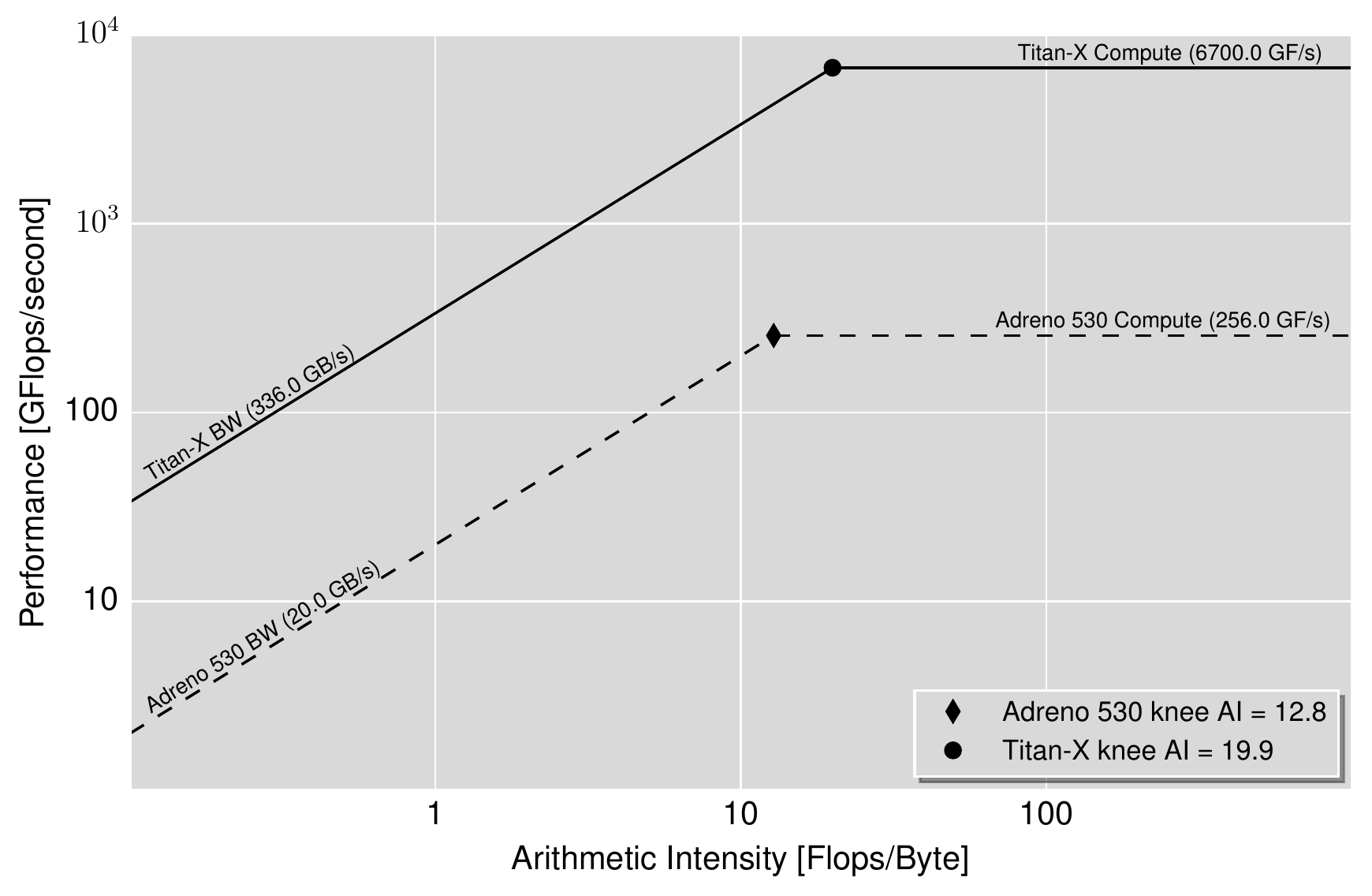}
  \caption{Roofline plot of peak compute rate and off-chip bandwidth for platforms used in this work.}
  \label{fig:roofline}
\end{figure}
For the two platforms we consider in this work, Figure~\ref{fig:roofline} shows the basic roofline curves formed by considering peak compute rate and off-chip memory bandwidth.
The knee Arithmetic Intensity (AI) for each platform is marked; note that although the two platforms have very different absolute performance, their knee AIs are similar.
Peak values are taken from documentation in the case of the Nvidia hardware, but (due to lack of documentation) are approximated using a microbenchmark (clpeak~\cite{clpeak}) for the Qualcomm hardware.

\section{Warmup: Code Generation for SGEMM}
\label{sec:sgemm}
\vsp
\begin{figure}[ht]
  \centering
  \includegraphics[width=0.8\linewidth]{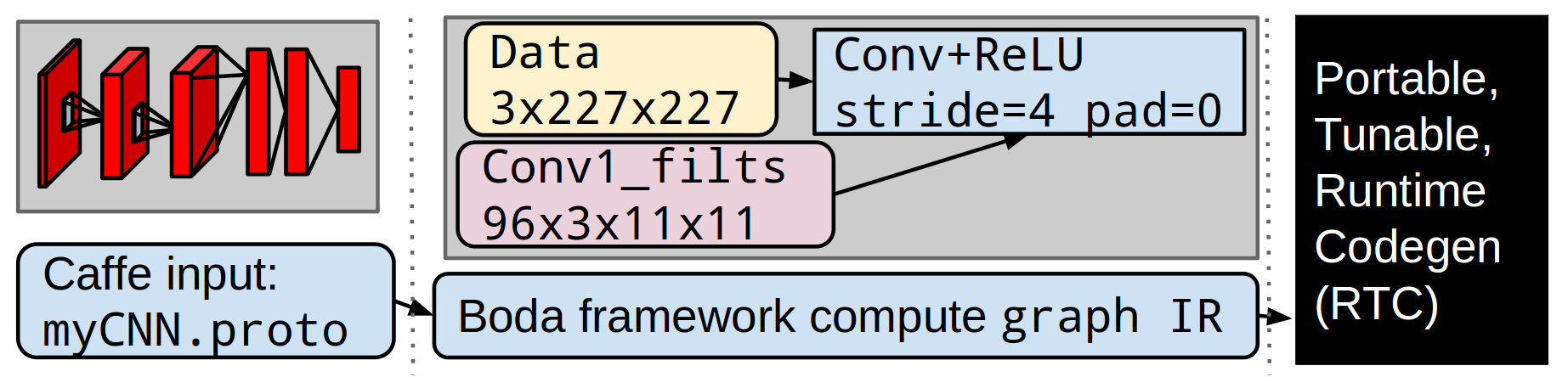}
  \caption{\boda~flow: CNN $\rightarrow$ compute graph $\rightarrow$ code generation.}
  \label{fig:boda-flow}
\end{figure}
At a high level, our framework employs a code generation flow summarized in Figure~\ref{fig:boda-flow}.
To explain the details of how this flow works in practice, we start with the application of our approach to a simple example: matrix-matrix multiplication.
As discussed earlier, early approaches to CNN convolutions employed matrix-matrix multiplication as the key computational primitive.
This allowed the usage of high-quality vendor provided BLAS libraries (such as cuBLAS from Nvidia) for the bulk of the computation.
In particular, one key BLAS function, single precision matrix-matrix multiply, or SGEMM, implements the core computation.
Also, following the Caffe naming convention, an auxiliary function $inmat = im2col(in)$ provides the conversion from the ND-Array $in$ to the 2D-matrix-of-input-slices $inmat$.
Note that $filtsmat$ is typically simply a reshape of $filts$.
As mentioned earlier, the expansive nature of $im2col()$ can be problematic.
In particular, for large batches of images, the size of $inmat$ (for a single layer) may exceed GPU memory.
To avoid this, batches may be handled with separate per-image calls to $im2col()$ and $SGEMM()$.
SGEMM is a simple and well studied function across many hardware architectures.
The main required tasks for achieving an efficient SGEMM implementation for a given hardware target are as follows:
\begin{enumerate}
\item Arranging accesses to storage to make best use of system's communication potential.
\item Achieving sufficient data reuse at each level of the system's storage hierarchy to avoid being bandwidth limited.
\item Ensuring computational units are continually active; (2) is necessary but not sufficient for this.
\item Repeating (1), (2), and (3) for all interesting input sizes.
\end{enumerate}

Unfortunately, these goals are often both interrelated and in conflict with each other.
For example, it is often the case that accessing storage contiguously, or in certain patterns, achieves higher bandwidth than others.
Thus, there may be a tradeoff between achieving good communication bandwidth and reading the best set of data for reuse at the next level.
In general, (1) and (2) directly balance each other.
So, if a factor of N additional reuse can be achieved at anything less than a factor of N cost in bandwidth, it is favorable to do so.
The primary goal is (3); if the maximum compute rate can be achieved, other concerns are secondary.
The input sizes for SGEMM are expressed as $M$, $K$, and $N$, all scalars, where if $c=SGEMM(a,b)$, then $c$ is an $M \dx N$ matrix, $a$ is $M \dx K$, and $b$ is $K \dx N$.
The general-case AI calculation for SGEMM is $AI = \frac{2MNK}{4(MN+MK+KN)}$.
Typically, it is difficult to achieve good efficiency across a range of sizes and hardware targets without some form of \textit{metaprogramming}.
We define metaprogramming as the collection of techniques where, instead of directly writing code, some higher level facility is used to create the desired final code.
C Macros, C++ templates, and ad-hoc code generation are all examples of metaprogramming, and all offer similar key benefits:
\begin{itemize}
\item In general, values that are known ahead of time can be constant in the final code without (potentially repeatedly) hard-coding them at the source level.
\item Loops with static bounds are easier to unroll and/or require less instructions or registers to implement.
\item Offsets, strides, scales, and other values can often can statically combined and/or used as immediates to reduce register usage and instruction count.
\item Conditionals depending on known-constant values can be eliminated.
\item Many variants of a single version of an algorithm can be generated simply by varying parameters at the meta level.
\item Repetitive sections of code can be generated, rather than manually written. This is particularly important when simpler techniques such as compiler-driven loop unrolling are insufficient or cumbersome.
\end{itemize}

The usage of metaprogramming for SGEMM for GPUs seems to be commonplace; as far the authors are aware, it is used to varying degrees by all modern, efficient GPU BLAS libraries including cuBLAS~\cite{cuBLAS}, MAGMA~\cite{magma-autotuning}, and clBLAS~\cite{clblas}.

Our framework uses a combination of string-replacement templates, specific support for known-size ND-Array access, and ad-hoc unrestricted programmatic code generation.
This places it toward the more flexible/extreme end of the space of metaprogramming techniques.
Although metaprogramming is inherently complex, we nonetheless attempt to achieve a good balance between flexibility, power, simplicity, and ease of use.
To implement SGEMM in OpenCL for GPU targets, we employ the following standard techniques:
\begin{itemize}
\item Register Tiling~\cite{iandola2013communication}
\item Explicit Local Memory Blocking
\item Inner Loop Unrolling
\end{itemize}
Listing~\ref{lst:sgemm} shows pseudocode for our SGEMM template.
\lstset{basicstyle=\ttfamily\scriptsize}
\begin{lstlisting}[float=ht,caption={SGEMM code template},label={lst:sgemm}]
void SGEMM( nda M:K a, nda N:K bt, nda M:N c ) {
  // dims work Mg:Ng:Mb:Nb:Kb:Mt:Nt // blocking values
  local a_lm[%(work_Kb_dim)*%(work_Mb_dim)*%(work_Mt_dim)];
  local b_lm[%(work_Kb_dim)*%(work_Nb_dim)*%(work_Nt_dim)];
  // per-thread tile of output to compute, stored in registers
  float c_t[%(work_Mt_dim)*%(work_Nt_dim)] = {0}; 
  float a_r[%(work_Mt_dim)]; 
  float b_r[%(work_Nt_dim)];
  for( int32_t k = 0; k < %(a_K_dim); k += %(work_Kb_dim) ) {
    BARRIER_SYNC;
    // workgroup-wide load of local memory for this iteration
    %(lm_loads); 
    BARRIER_SYNC;
    for( uint32_t subk = 0; subk < %(work_Kb_dim); ++subk ) {
      %(loads); // load per-thread slice of a, b into a_r, b_r
      %(fmas); // perform fused multiply-adds
    }
  }
 // iterate over rows of the Mt * Nt registers in c_t
  for( int32_t m = 0; m < %(work_Mt_dim); ++m ) { 
    %(transpose_c_t_row); // transpose row of c_t into b_r
    %(store_c_t_row); // store transposed row of c_t to c
  }
}
\end{lstlisting}
\ifdefined\inmaindoc
\else
\documentclass{article}
\usepackage{booktabs}       
\begin{document}
\newcommand{\dx}{{\times}}
\fi
\begin{table*}[t] 
\caption{SGEMM operation speed and efficiency: cuBLAS vs. Boda}
\centering
\begin{tabular}{llllllllc}
\hline
    \multicolumn{1}{l}{Size} & \multicolumn{3}{l}{Communication/Compute} & 
    \multicolumn{2}{l}{cuBLAS Performance} & \multicolumn{2}{l}{Boda Performance} \\
    \cmidrule(r){1-1} \cmidrule(r){2-4}  \cmidrule(r){5-6} \cmidrule{7-8}
    MKN & Bytes & FLOPs & F/B & Runtime & GF/s & Runtime & GF/s & Speedup \\
    \hline
 $ 128 $ & 197KB & 4.19MF & 21.3  & 49.3us & 85.0GF/s  & 36.7us & 114GF/s  & 1.34x \\ 
 $ 256 $ & 786KB & 33.6MF & 42.7  & 49.2us & 681GF/s  & 54.2us & 619GF/s  & 0.91x \\ 
 $ 384 $ & 1.77MB & 113MF & 64.0  & 63.3us & 1.79TF/s  & 80.1us & 1.41TF/s  & 0.79x \\ 
 $ 512 $ & 3.15MB & 268MF & 85.3  & 107us & 2.51TF/s  & 120us & 2.24TF/s  & 0.89x \\ 
 $ 768 $ & 7.08MB & 906MF & 128  & 202us & 4.49TF/s  & 255us & 3.55TF/s  & 0.79x \\ 
 $ 1024 $ & 12.6MB & 2.15GF & 171  & 541us & 3.97TF/s  & 602us & 3.57TF/s  & 0.90x \\ 
 $ 1536 $ & 28.3MB & 7.25GF & 256  & 1.39ms & 5.22TF/s  & 1.63ms & 4.44TF/s  & 0.85x \\ 
 $ 2048 $ & 50.3MB & 17.2GF & 341  & 3.56ms & 4.83TF/s  & 4.08ms & 4.21TF/s  & 0.87x \\ 
  \hline
  \label{tab:sgemm-titanX-cuBLAS-boda-comp}
\end{tabular}
\end{table*}

\ifdefined\inmaindoc
\else
\end{document}
\fi

\begin{figure}[ht]
  \centering
  \includegraphics[width=0.8\linewidth]{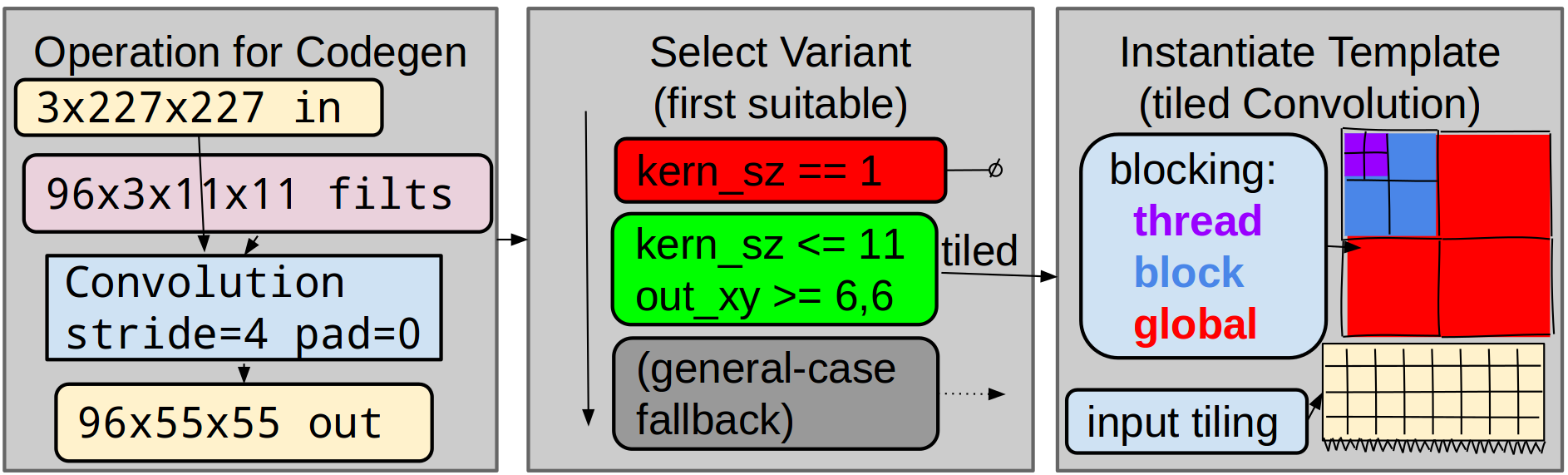}
  \caption{\boda~codegen: CNN op $\rightarrow$ Variant $\rightarrow$ blocking and GPU code.}
  \label{fig:boda-codegen}
\end{figure}

For a given input size and hardware platform, the first main task of the code generation is to determine a good blocking strategy.
This can be accomplished with a combination of heuristic calculations, manual parameter tuning, or automated tuning.
In this work, we use only the first two approaches; use of automated tuning is a good subject for future work.
The general flow from operation (SGEMM or convolution) to blocking constants is illustrated in Figure~\ref{fig:boda-codegen}.
The main blocking parameters we must choose are:
\begin{itemize}
\item $Mt, Nt$: The number of output elements computed per thread in the M and N dimensions. Typically in the range $[1,8]$. Note that the kernel will require at least $Mt \dx Nt$ registers per thread.
\item $Kb$: The inner loop unrolling factor. Typically in the range $[1,8]$. If too small, loop overhead becomes excessive. If too large, Instruction or Local memory will be insufficient, as both scale linearly with $Kb$.
\item $Mb,Nb$: Determines the total number of output elements computed per workgroup. Valid and/or desirable values vary by hardware architecture, but often $Mb \dx Nb \equiv threads\_per\_workgroup$ should be in the range $[32,256]$. When $Mb \approx Nb$, both Local memory usage and the Global/Local memory reuse factor scale linearly as $Mb$ (or $Nb$).
\item $Mg,Ng$: Based on $M,N$ and the prior choices of $Mt,Nt,Mb,Nb$, we have $Mg = \ceil{M/Mb/Mt}$ and $Ng = \ceil{N/Nb/Nt}$. $Mg \dx Ng$ workgroups will be needed to compute the final result. For many targets, the total number of workgroups must be above a threshold to saturate all computational elements in the device. Small values may be subject to performance limitations if they are not a multiple of some particular constant.
\end{itemize}

After determining the blocking constants, the code generator must emit code for various blocks:
\begin{itemize}
\item \%(lm\_loads): workgroup-wide cooperative global memory $\rightarrow$ local memory loads
\item \%(loads):$Mt + Nt$ per-thread local memory $\rightarrow$ register loads into $a\_r[m], b\_r[n]$
\item \%(fmas): $Mt \dx Nt$ in-register multiply-adds: $c\_t[m][n] += a\_r[m] * b\_r[n]$
\item \%(transpose\_c\_t\_row), \%(store\_c\_t\_row): transpose and write per-thread outputs into $c$
\end{itemize}

The general challenges when emitting these blocks are to minimize conditionals and choose particular constructs that can execute efficiently.
Careful data layout at the global, local, and register levels may be required, potentially with additional code to reorganize, shuffle, or transpose data at each level.
With a few days of effort, reasonable SGEMM performance on an Nvidia Titan-X GPU was achieved.
See Table~\ref{tab:sgemm-titanX-cuBLAS-boda-comp} for a comparison of this template's performance with that of the SGEMM from Nvidia's highly-tuned cuBLAS library.
In short, with moderate effort our framework achieved $\sim$ 80\% of the performance of the vendor library, albeit only for a few simple cases.

\ifdefined\inmaindoc
\else
\documentclass{article}
\usepackage{booktabs}       
\begin{document}
\newcommand{\dx}{{\times}}
\fi

\begin{table*}[t] 
\caption{SGEMM operation speed and efficiency: QSML (CPU; no vendor-provided GPU support) vs. Boda (GPU)}
\centering
\begin{tabular}{llllllllc}
\hline
    \multicolumn{1}{l}{Size} & \multicolumn{3}{l}{Communication/Compute} & 
    \multicolumn{2}{l}{QSML Performance} & \multicolumn{2}{l}{Boda Performance} \\
    \cmidrule(r){1-1} \cmidrule(r){2-4}  \cmidrule(r){5-6} \cmidrule{7-8}
    MKN & Bytes & FLOPs & F/B & Runtime & GF/s & Runtime & GF/s & Speedup \\
    \hline

$128$ & 197KB & 4.19MF & 21.3 & 560us & 7.5GF/s & 264us & 16GF/s & 2.1x \\ 
$256$ & 786KB & 33.6MF & 42.7 & 1.7ms & 20GF/s & 507us & 66GF/s & 3.3x \\ 
$384$ & 1.77MB & 113MF & 64.0 & 3.8ms & 30GF/s & 2.1ms & 54GF/s & 1.8x \\ 
$512$ & 3.15MB & 268MF & 85.3 & 8.0ms & 34GF/s & 3.6ms & 74GF/s & 2.2x \\ 
$768$ & 7.08MB & 906MF & 128 & 23ms & 39GF/s & 11ms & 78GF/s & 1.9x \\ 
$1024$ & 12.6MB & 2.15GF & 171 & 54ms & 40GF/s & 27ms & 80GF/s & 2.0x \\ 
$1536$ & 28.3MB & 7.25GF & 256 & 175ms & 41GF/s & 89ms & 81GF/s & 1.9x \\ 
$2048$ & 50.3MB & 17.2GF & 341 & 401ms & 42GF/s & 223ms & 77GF/s & 1.8x \\ 
  \hline
  \label{tab:sgemm-SD820-QBLAS-boda-cmp}
\end{tabular}
\end{table*}

\ifdefined\inmaindoc
\else
\end{document}
\fi

Next, we turn our attention to our main focus: our new hardware target, the Snapdragon 820 (SD820).
The first key thing we learned about this platform is that manual vectorization of load and stores yields an improvement of 2x or more in load/store bandwidth.
Further, at least for SGEMM on the SD820 platform, we find that it is not generally profitable to explicitly move data from global to local memory.
Instead, we apply our usual work-blocking strategy, but simply omit the loads and stores to local memory, and read/write global memory directly.
Due to the access pattern of the blocking, the hardware cache appears to provide data reuse similar to that of using local memory explicitly, and we avoid the overhead of both the local memory accesses and related synchronization overheads.
However, note that the resulting bandwidth amplification is limited, perhaps due to overall limited bandwidth from cache/local-memory.
Unfortunately, the SD820 development platform does not provide sufficient profiling information, hardware/ISA documentation, or other tools (such as a disassembler) to perform a more in depth performance analysis.
As a result, days were required to improve results through blind experimentation, guesswork, and tuning.
In the end, we adapted our approach for the SD820 using the following techniques:
\begin{itemize}
\item Manual vectorization of loads/stores
\item Local-memory based output buffers (which seems to a compiler optimization triggered by high register use)
\item Direct global memory access with cache Blocking (i.e. don't attempt to use local memory explicitly)
\end{itemize}
The resulting speed of our SGEMM on the SD820 platform is 2X that of the SGEMM from the vendor provided QSML~\cite{qsml} library; see Table~\ref{tab:sgemm-SD820-QBLAS-boda-cmp}.
Note, however, that there is currently no vendor provided GPU-based SGEMM, so this comparison is against code running on the CPU portion of the SD820 SoC.
Currently, it is not clear if any other OpenCL BLAS libraries can target the SD820 platform without some significant additional efforts.
Thus, research of, profiling of, and comparison against other OpenCL-based BLAS libraries is a good topic for future work.

\section{Code Generation for CNN Convolutions}
\label{sec:gencnnconv}
\vsp

As discussed earlier, the BLAS-based or $im2col()$/$SGEMM()$ approach to performing convolutions has various limitations.
Now, we turn our attention to using our framework to generate functions that directly perform convolutions.
The first variant we will discuss is a simple fusion of $im2col()$ and $SGEMM()$.
We apply the same basic techniques as in the $SGEMM()$ discussion above, but we fold the behavior of $im2col()$ into the \%(lm\_loads) code block.
Or, in other words, we only \textit{implicitly} create the matrix $inmat$; we just read the correct elements from $in$ as needed.
While this approach is simple, and avoids the memory overhead of creating $inmat$ explicitly, it makes task (1) much more difficult.
In particular, both the overhead of additional indexing logic and the resultant poor access patterns reading global memory can make this variant less efficient than $im2col() + SGEMM()$.
However, it provides a starting place for further explorations, and can function as a fallback method for convolutions not handled by more specialized variants, especially for cases with large kernel sizes where the overhead of $im2col()$ is larger.
We term this the \textit{implicit-SGEMM} or \verb|conv| variant.
The next variant we consider exploits the common case where the convolution kernel size $KSZ$ is 1.
In this case, various simplifications are possible, and it is relatively easy to use a transformation function over $in$ to ensure a good global memory access pattern.
Note that, for $KSZ=1$ convolutions, per-image $im2col()$ is the identity function; thus the \verb|k1conv| variant is quite similar to SGEMM.
The final variant we consider, termed \verb|tconv| (tiled convolution), is targeted at the commonly occurring cases of kernel sizes in the range $[2,\sim11]$, with reasonable widths for $in$ (perhaps in the range $[KSZ*5,KSZ*50]$).
In this case, we can perform some additional optimizations:
\begin{itemize}
\item We can fully unroll over the X dimension of the kernel. This uses significant but limited extra local memory and registers, but allows sharing of $in$ row data in registers across unrollings of the inner loop.
\item We can load entire X/Y \textit{tiles} of $in$ at the workgroup level. Even for kernels as small as 2x2, this vastly reduces the amount of data that must be loaded from global memory for $in$. The reduction is a factor of $(KSZ/stride)^2$; this is naturally the same factor by which $im2col()$ is expansive.
\end{itemize}
Again, an input transformation must be applied to $in$ to help simplify indexing logic and improve memory access patterns.

For benchmarking, we consider a range of convolutions drawn from three common CNNs: AlexNet~\cite{alexnet}, NiN~\cite{Lin2013NiN}, and GoogLeNetV1~\cite{googlenet}.
For each network, we consider batch sizes $B$ of 1, 5, and 20.
We then gather all the unique convolution operations, of which there are $\sim180$.
While some operations are duplicated within some networks, and the mixing together of operations from different batch sizes is perhaps not ideal, this set of operations represents a reasonable set over which low total runtime (summed over all operations) is desired.
That is, absolute efficiency is generally less important than absolute runtime, and total runtime tends to be dominated by the larger convolutions.
That said, high efficiency across a broad range of problems sizes is still desirable.
Each particular network, batch size, and overall use case requires some particular subset of convolutions.
As with SGEMM, we initially compare our convolutions against Nvidia's cuDNN library on an Nvidia Titan-X.
Due to space limitations, we do not present the full results of that experiment here, but only summarize them.
In brief, as with SGEMM, we achieve reasonable performance, albeit with a few weeks of effort rather than a few days.
Compared to SGEMM, tuning convolutions took more time for various reasons.
This was the both the author's first experience at implementing high efficiency code on Maxwell GPUs as well as the author's first attempt at implementing convolutions; both of these incurred a substantial learning curve penalty.
Beyond that, the design space for convolution seems to be more complex and varied than that of SGEMM; while this does offer more potential for optimization, it certainly also increases development time to the degree one attempts to explore and exploit the space.
Note that of the $\sim180$ convolutions, almost all are handled by either the \verb|k1conv| or \verb|tconv| variant.
Only a few cases fall though to the \verb|conv| variant.
Both \verb|k1conv| and \verb|tconv| provide a 2x or more speedup over \verb|conv|.
Now, we turn to our main focus in this work of targeting the SD820 platform.
As with the SGEMM case, our main task is to manually vectorize loads and stores.
Additionally, as with SGEMM, we avoid the explicit use of local memory, and instead rely on cache for global memory bandwidth amplification.
So far, we have only implemented two new variants for the SD820 platform: \verb|conv_simd| and \verb|k1conv_simd|, which are manually load/store vectorized versions of their non-simd counterparts.
As with the Nvidia case, the \verb|k1conv_simd| variant provides significant speedup over the fallback \verb|conv_simd| variant.
Detailed speed results are presented for the SD820 platform in the next section.
Given the knowledge gained from implementing SGEMM on the SD820, it took only a few days to create these new variants.
As with SGEMM, progress was hindered due to lack of documentation and tools.
Such things are undoubtedly available inside of corporations, and we hope the potential for higher performance CNN implementations will encourage vendors to make them available to programmers.

\section{Results}
\label{sec:results}
\vsp

\begin{figure}[ht]
  \centering
  \includegraphics[width=.95\linewidth]{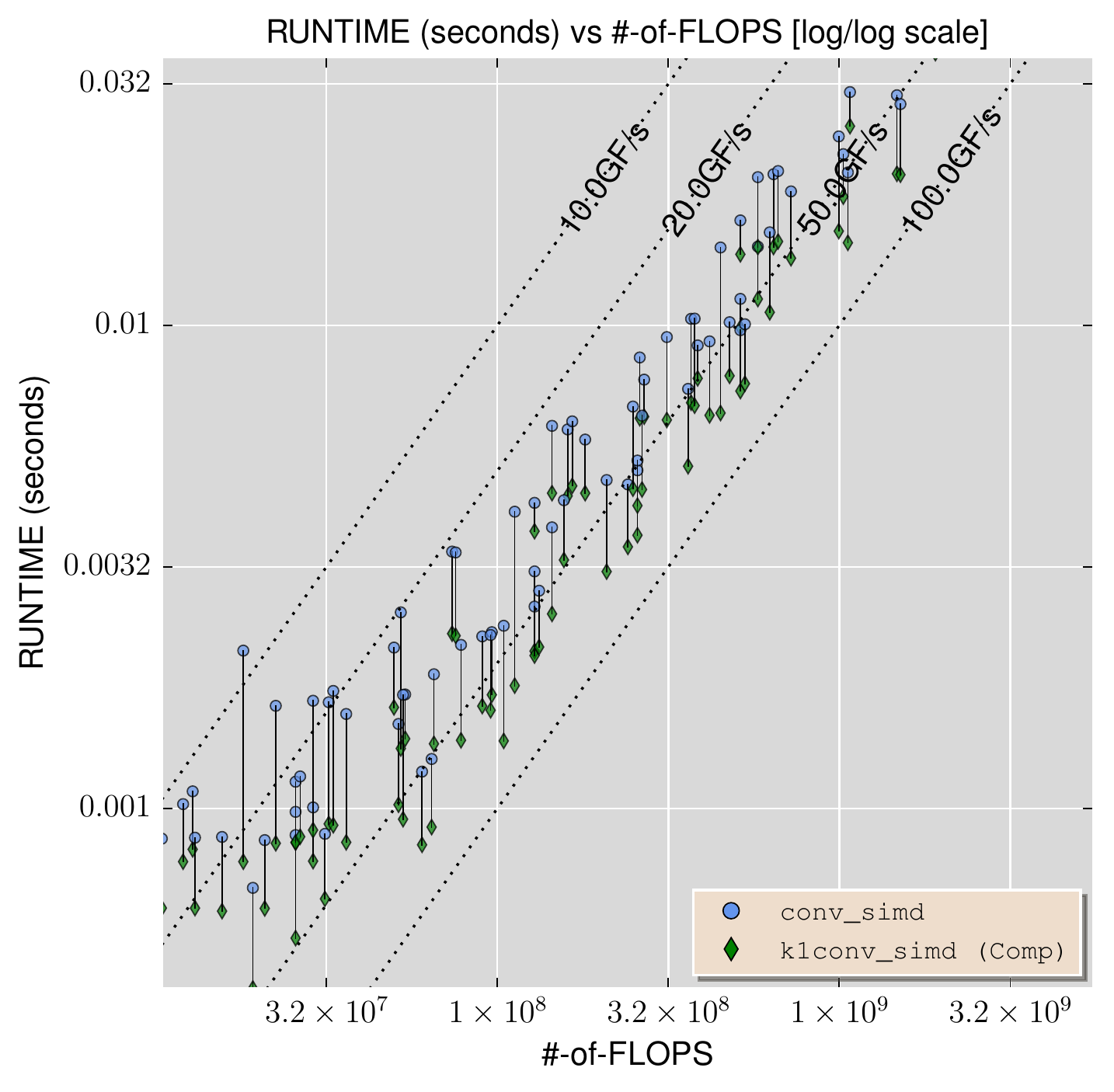}
  \caption{Speedup of k1conv\_simd over conv\_simd on SD820 platform.}
  \label{fig:k1conv}
\end{figure}

\begin{figure}[ht]
  \centering
  \includegraphics[width=0.9\linewidth]{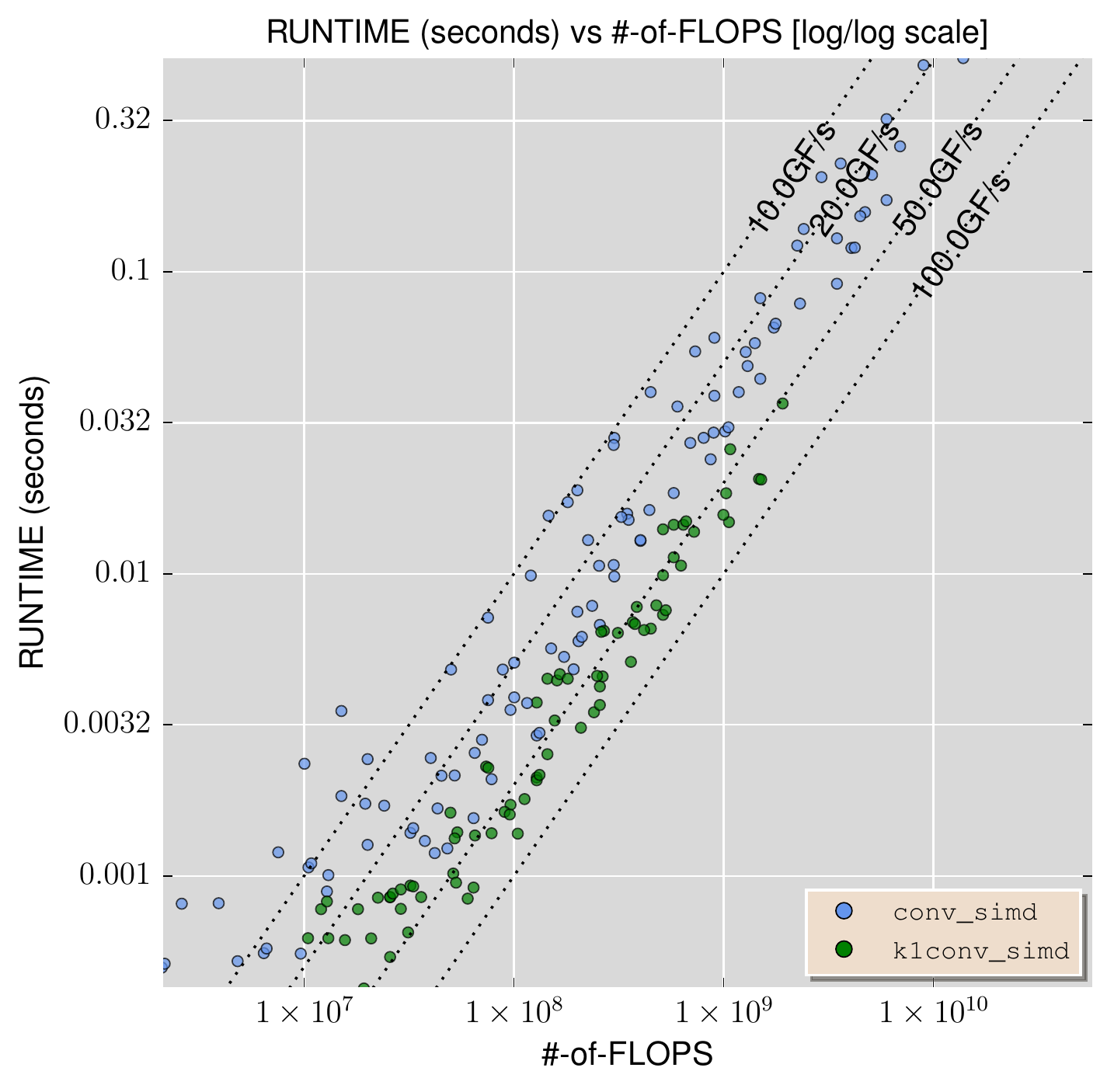}
  \caption{Per-convolution-size speed on SD820 platform.}
  \label{fig:SD820-results}
\end{figure}

\begin{figure}[ht]
  \centering
  \includegraphics[width=0.9\linewidth]{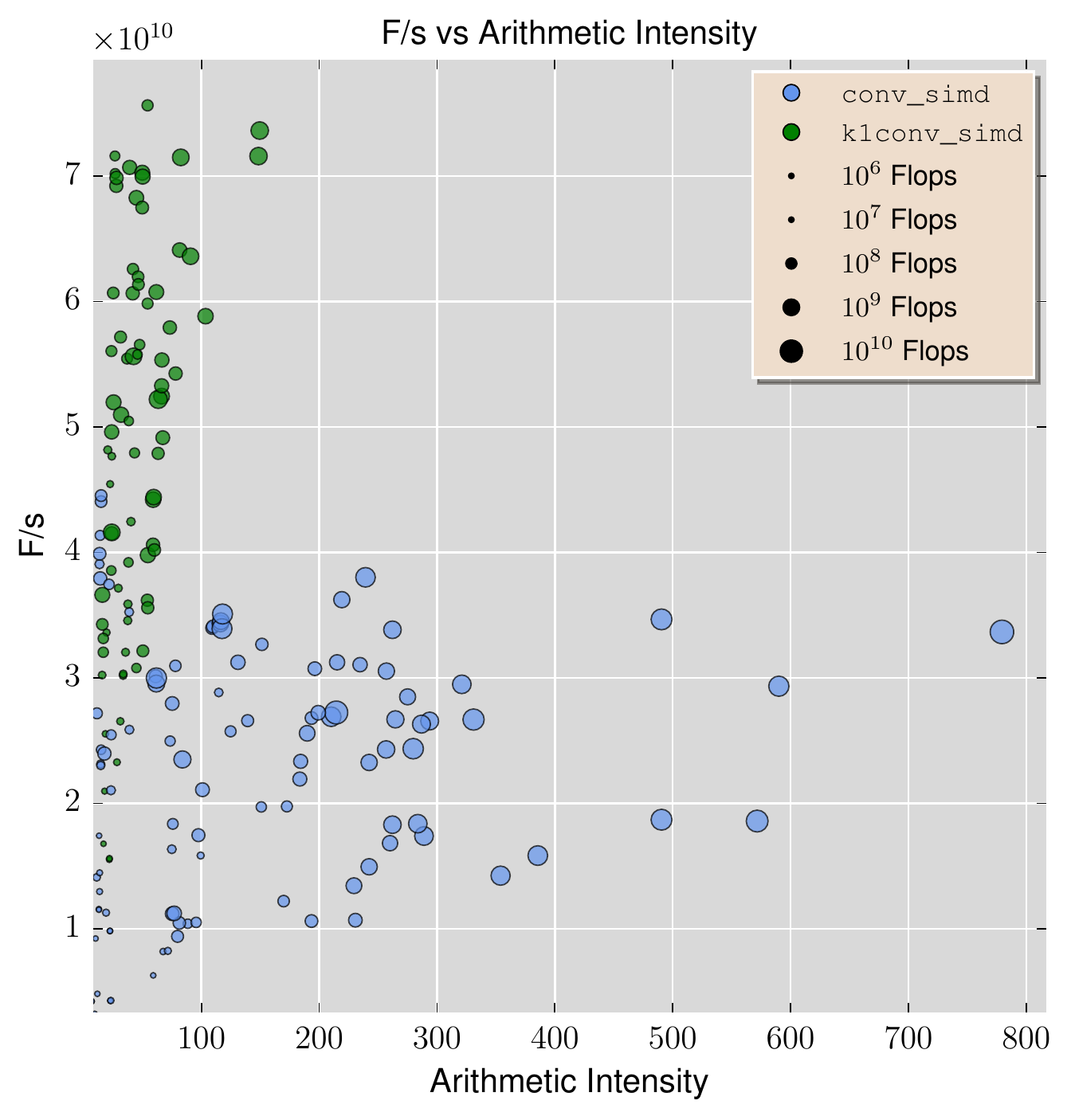}
  \caption{Same data as Fig~\ref{fig:SD820-results}, plotted as speed vs. arithmetic intensity. }
  \label{fig:SD820-results-ai}
\end{figure}

Here, for the SD820 platform, we show the speed of the OpenCL code generated by our framework for our benchmark set of convolutions.
In Figure~\ref{fig:k1conv}, we show the benefit of the \verb|k1conv_simd| variant for the convolutions to which it can be applied: those with size 1 kernels.
In Figures~\ref{fig:SD820-results} and~\ref{fig:SD820-results-ai}, we show the absolute performance of our framework's generated code for each benchmark convolution.
As per the graph legend, for each convolution, we indicate which variant was selected.
Inspecting the results, it can be seen that there are many convolutions with high arithmetic intensity (AI) that perform worse than those with lower AI.
This is due to the fact that the higher performing cases are using the higher-efficiency \verb|k1conv_simd| variant.
Based on our experience with the Nvidia platform, we predict that implementing a \verb|tconv_simd| variant for the SD820 platform will greatly improve the performance of most of the cases currently using the fallback \verb|conv_simd| variant.
Given the lack of a vendor CNN library or other libraries to directly compare against, it is difficult to know how close our performance is to optimal.
From the SD820 roofline curve in Figure~\ref{fig:roofline}, we know there is significant headroom over our results in terms of peak compute performance.
While our limited knowledge of the SD820's on-chip memory subsystems makes a determination difficult, in many cases it seems likely we are limited by cache and/or global memory bandwidth.
Thus, usage of smaller data types for storage (e.g. half-precision floats) and/or using hardware support for texture access are natural candidates to achieve additional improvements.

\section{Conclusions}
\label{sec:conclusions}
\vsp

Convolutional Neural Nets are of growing importance in a broad range of applications, and particularly in computer vision systems used in self-driving cars, medical imaging, and a variety of consumer-facing applications such as face identification in social media.
As both the research and development communities continue to grow, interest in productive deployment of CNNs across many platforms and application domains will only increase;
however, the number of programmers capable of efficiently implementing CNN operations is very limited, and the keys for efficient implementations are not widely known.
As a result, support for high efficiency CNN calculation is currently limited to only a few hardware platforms.
In the Boda framework described in this paper we have aspired to bridge the performance-portability gap for the key CNN operations and to bring such operations to a more even footing across various hardware platforms when compared with existing high efficiency approaches.
We have demonstrated our approach with a case study of tuning CNN deployment computations for the Snapdragon 820 mobile computing platform.
By offering competitive performance and superior portability, we feel this work will positively impact the ability of the research and development communities to experiment with the deployment of CNNs across a wide range of platforms for an ever-broadening range of applications.


\ifdefined\anon
\else
\subsubsection*{Acknowledgments}
Research partially funded by DARPA Award Number HR0011-12-2-0016, the Berkeley Deep Drive (BDD) Industry Consortium, and ASPIRE industrial sponsors and affiliates Intel, Google, Hewlett-Packard, Huawei, LGE, Nvidia, Nokia, Oracle, and Samsung.
\fi

{
\small
\bibliographystyle{IEEEtran}
\bibliography{bibliography}
}

\end{document}